\documentclass[12pt,a4paper]{article}
\usepackage{graphicx}
\usepackage{subeqnarray}
\begin{document}
%
%
%
%
\def\astrobj#1{#1}
\newenvironment{lefteqnarray}{\arraycolsep=0pt\begin{eqnarray}}
{\end{eqnarray}\protect\aftergroup\ignorespaces}
\newenvironment{lefteqnarray*}{\arraycolsep=0pt\begin{eqnarray*}}
{\end{eqnarray*}\protect\aftergroup\ignorespaces}
\newenvironment{leftsubeqnarray}{\arraycolsep=0pt\begin{subeqnarray}}
{\end{subeqnarray}\protect\aftergroup\ignorespaces}
\newcommand{\diff}{{\rm\,d}}
\newcommand{\pprime}{{\prime\prime}}
\newcommand{\szeta}{\mskip 3mu /\mskip-10mu \zeta}
\newcommand{\FC}{\mskip 0mu {\rm F}\mskip-10mu{\rm C}}
\newcommand{\appleq}{\stackrel{<}{\sim}}
\newcommand{\appgeq}{\stackrel{>}{\sim}}
\newcommand{\legr}{\stackrel{<}{>}}
\newcommand{\grle}{\stackrel{>}{<}}
\newcommand{\Int}{\mathop{\rm Int}\nolimits}
\newcommand{\Nint}{\mathop{\rm Nint}\nolimits}
\newcommand{\range}{{\rm -}}
\newcommand{\displayfrac}[2]{\frac{\displaystyle #1}{\displaystyle #2}}
\def\astrobj#1{#1}
%
\title{
Simple MCBR models of chemical evolution: \\
an application to the thin and the thick disk}
\author{{R.~Caimmi}\footnote{
{\it Physics and Astronomy Department, Padua Univ., Vicolo Osservatorio 3/2,
I-35122 Padova, Italy}
email: roberto.caimmi@unipd.it~~~
fax: 39-049-8278212}
\phantom{agga}}
%
%
\maketitle
\begin{quotation}
\section*{}
\begin{Large}
\begin{center}

Abstract

\end{center}
\end{Large}
\begin{small}

\noindent\noindent
Simple multistage closed-(box+reservoir) (MCBR) models of chemical
evolution, formulated in an earlier attempt, are
extended to the limit of dominant gas inflow or outflow with respect to gas
locked up into long-lived stars and remnants.   For an assigned empirical
differential
oxygen abundance distribution (EDOD), which can be linearly fitted, a family
of theoretical differential oxygen abundance distribution (TDOD) curves is
built up with the following prescriptions: (i) the initial and the ending
point of the linear fit are in common to all curves; (ii) the flow parameter,
$\kappa$, ranges from an extremum point to $\mp\infty$, where negative and
positive $\kappa$ correspond to inflow and outflow, respectively; (iii) the
cut parameter, $\zeta_{\rm O}$, ranges from an extremum point (which cannot be
negative) to the limit, $(\zeta_{\rm O})_\infty$, related to $\vert\kappa\vert
\to+\infty$.   For curves with increasing $\zeta_{\rm O}$, the gas mass
fraction locked up into long-lived stars and remnants is found
to attain a maximum and then decrease towards zero as $\vert\kappa\vert\to
+\infty$, while the remaining parameters
show a monotonic trend.   The theoretical integral oxygen abundance
distribution (TIOD) is also expressed.   An application is performed to the
EDOD deduced from two
different samples of disk stars, for both the thin and the thick disk.   The
constraints on formation and evolution are discussed in the light of the 
model.   The evolution is tentatively subdivided into four stages, namely:
assembling (A), formation (F), contraction (C), equilibrium (E).   The EDOD
related to any stage is fitted by all curves where $0\le\zeta_{\rm O}\le
(\zeta_{\rm O})_\infty$ for inflowing gas and $(\zeta_{\rm O})_\infty\le
\zeta_{\rm O}\le1.2$ for outflowing gas, with a single exception related to
the thin disk (A stage), where the range of fitting curves is restricted to
$0.35\le\zeta_{\rm O}\le(\zeta_{\rm O})_\infty$.   The F stage may safely be
described by a steady inflow regime $(\kappa=-1)$, implying a flat TDOD, in
agreement with the results of hydrodynamical simulations.   Finally, (1) the
change of fractional mass due to the extension of the linear fit to the EDOD,
towards both the (undetected) low-metallicity and high-metallicity tail, is
evaluated and (2) the idea
of a thick disk-thin disk collapse is discussed, in the light of the model.

\noindent
{\it keywords - 
galaxies: evolution - stars: formation; evolution.}
\end{small}
\end{quotation}

\section{Introduction} \label{intro}

The empirical metallicity distributions of long-lived
stars, belonging to different populations, constrain models for
the formation and the evolution of the Galaxy.  Simple
multistage closed-(box+reservoir) (MCBR) models formulated in an earlier
attempt (Caimmi 2011a, hereafter quoted as C11) make a useful tool in the
description of galactic chemical evolution.   In particular, MCBR models
allow for gas inflow or outflow with specified metal abundance ratio with
respect to the pre existing gas, but general conclusions can also be inferred
(C11).   The special case of steady inflow regime, where the
infalling gas exactly balances the net amount of gas turned into stars, finds
a counterpart in the results of hydrodynamical simulations, where quasi
equilibrium is attained between inflowing gas, outflowing gas and gas lost via
star formation (e.g., Finlator and Dav\'e 2008; Dav\'e et al. 2011a,b, 2012).

For different populations within the Galaxy, the empirical differential
oxygen abundance distribution (EDOD) shows the existence of three different
regimes of chemical evolution, where gas inflow is initially the dominant
process, followed by a nearly steady inflow regime, while gas outflow prevails
at late times (C11 and further references therein).   Understanding different
stages of Galactic (or sub Galactic) chemical evolution can be used to get
further insight on the evolution of important global properties of stellar and
gaseous component of galaxies.

The current paper is aimed to (i) extension and improvement of MCBR models and
(ii) application to the thin and the thick disk.   The extension of the model
is concerned with the limit of dominant inflowing or outflowing gas rate in
comparison to star formation rate and calculation of related quantities which,
for assigned input values, restricts the parameter space.   In addition, the
theoretical integral oxygen abundance distribution (TIOD) is explicitly
expressed.   The
improvement of the model consists in a different choice of the family of
fitting theoretical differential oxygen abundance distribution (TDOD) curves
with respect to an earlier attempt (C11), which can be used for any kind of
environment.

With regard to an assigned stage of evolution, the starting and
the ending point of the linear fit to the EDOD are in common to each curve of
the family, instead of the intercept and the ending point as in the parent
paper (C11).   The difference is negligible if the model is applied to
populations with initial oxygen abundance close to zero such as the halo
(C11), while it is consistent for populations with initial oxygen abundance
substantially larger than zero such as the thin and the thick disk, which are
dealt with in the current paper.

The EDOD for the thin and the thick disk is inferred from two different
samples in Section 2.   Extension and improvement of MCBR models is performed
in Section 3.   An application to the thin and the thick disk is shown in
Section 4.   The results are discussed in Section 5.   The conclusion is drawn
in Section 6.

\section{The EDOD for the thin and the thick disk} \label{EDOD}
\subsection{General remarks} \label{gerE}

The abundance distribution of a generic nuclide, Q, for star samples, is
binned in [Q/H]$=\log(n_{\rm Q}/n_{\rm H})-\log(n_{\rm Q}/n_{\rm H})_\odot$,
where $n_{\rm Q}$, $n_{\rm H}$, are number abundances.   On the other hand,
the comparison between the empirical differential metal abundance distribution
and its theoretical counterpart predicted by a model implies the knowledge of
the normalized mass abundance, $\phi_{\rm Q}=Z_{\rm Q}/(Z_{\rm Q})_\odot$.
It can be seen that $\log\phi_{\rm Q}=$[Q/H] to a good extent.   For further
details refer to an earlier attempt (Caimmi 2007).

In terms of mass abundance, the bin centre and the bin semiamplitude read:
\begin{lefteqnarray}
\label{eq:phic}
&& \phi_{\rm Q}=\frac12\{\exp_{10}{\rm[Q/H]}^++\exp_{10}{\rm[Q/H]}^-\}~~; \\
\label{eq:phia}
&& \Delta^\mp\phi_{\rm Q}=\frac12\{\exp_{10}{\rm[Q/H]}^+-\exp_{10}{\rm[Q/H]}^
-\}~~;
\end{lefteqnarray}
which implies a variable bin width in $\phi_{\rm Q}$ for a constant bin width
in [Q/H].

The differential metal abundance distribution has been dealt with in earlier
attempts in both non normalized (Pagel 1989; Malinie et al. 1993) and
normalized (Rocha-Pinto and Maciel 1996; Caimmi 2000, 2001a,b, 2007) form, the
last one to be used in the following.   The related bin centre and the bin
semiamplitude read:
\begin{lefteqnarray}
\label{eq:psie}
&& \psi_{\rm Q}=\log\frac{\Delta N}{N\Delta\phi_{\rm Q}}~~; \\
\label{eq:Dpsie}
&& \Delta^\mp\psi_{\rm Q}=\log\left[1\mp\frac{(\Delta N)^{1/2}}{\Delta N}
\right]~~;
\end{lefteqnarray}
where $\Delta N$ is the number of sample stars
within the metal abundance bin, $\Delta\phi_{\rm Q}=\Delta^-\phi_{\rm Q}+
\Delta^+\phi_{\rm Q}$, centered on $\phi_{\rm Q}$, $N$ is the total number of
sample stars, and the uncertainty on $\Delta N$ has been evaluated from
Poissonian errors as $\sigma_{\Delta N}=(\Delta N)^{1/2}$ (e.g., Ryan and
Norris 1991).    It is worth
noticing the semiamplitudes, $\Delta^\mp\psi_{\rm Q}$, are different and, in
particular, $\Delta^-\psi_{\rm Q}\to-\infty$ for $\Delta N=1$, which implies
caution has to be used with bins containing a single star.   For further
details refer to earlier attempts (Caimmi 2001b, 2007).

From this point on, attention shall be restricted only to oxygen, Q=O, where
$\phi_{\rm O}=\phi$ and $\psi_{\rm O}=\psi$ shall be used to simplify the
notation.   On the other hand, the results of the current paper hold for any
primary element which is mainly synthesized within type II supernova
progenitors.

\subsection{The data} \label{data}

Oxygen abundance determination in stellar atmospheres is intrinsically
difficult, in that different methods yield different results (e.g., Ramirez et
al. 2007; Fabbian et al. 2009) and no general consensus still exists.   For
this reason, the population of available samples does not exceed a few
hundredths at most.   For further details refer to earlier attempts (Caimmi
2007, 2008; Caimmi and Milanese 2009; C11).

With regard to the disk, the
data shall be inferred from samples studied in earlier attempts (Ramirez et
al. 2007; Petigura and Marcy 2011), where oxygen
abundance has been determined both in presence and in absence of the local
thermodynamical approximation (LTE) in the former case and only in presence of
LTE in the latter.   The samples extracted from the above mentioned parent
papers are listed in Table \ref{t:samp}.
\begin{table}
\caption{Samples used for determining the empirical differential oxygen
abundance distribution (EDOD) of the thin and the thick disk.   Sample
denomination relates to the parent sample (Ra07 - Ramirez et al. 2007; PM11 -
Petigura and Marcy 2011) and to the population [k - thick disk (TK); n - thin
disk (TN); stars uncertain if TK or TN are denoted as UN].   Presence or
absence of the LTE approximation for oxygen abundance determination is
denoted as yes or no, respectively.  The total number of sample stars is $N$.}
\label{t:samp}
\begin{center}
\begin{tabular}{llll} \hline
\multicolumn{1}{l}{sample} &
\multicolumn{1}{c}{population} &
\multicolumn{1}{l}{$N$} &
\multicolumn{1}{c}{LTE} \\
\hline
      &         &         &     \\
Ra07k & TK      & 133     & no  \\
Ra07n & TN      & 310     & no  \\
PM11k & TK + UN & 14 + 20 & yes \\
PM11n & TN      & 624     & yes \\
\hline                            
\end{tabular}                     
\end{center}                      
\end{table}                       

With regard to Ra07 samples, the EDOD has been inferred in an earlier attempt
(Caimmi and Milanese 2009) both in presence and in absence of the LTE
approximation and only the latter case shall be considered in the
current investigation.

The EDOD inferred from PM11 samples is presented in Table \ref{t:psph}.
\begin{table}
\caption{The empirical, differential
oxygen abundance distribution (EDOD) in the
thin and thick disk,
deduced from the PM11 sample
($N=658$).
The error on the generic bin height
has been estimated from the Poissonian error.
Disk stars belonging to uncertain population $(N=20)$
have been included within thick disk stars $(N=14)$
to get larger population $(N=34)$.   The bin width
in [O/H]$=\log\phi$ is $\mp0.05$. Label captions:
TN - thin disk; TK - thick disk; UN - uncertain if
belonging to thin or thick disk.   See text
for further details.}
\label{t:psph}
\begin{center}
\begin{tabular}{lllrrr} \hline
\multicolumn{1}{c|}{}
&\multicolumn{1}{c|}{TN}
&\multicolumn{1}{c|}{TK + UN}
&\multicolumn{3}{c}{$\Delta N$} \\
\hline\noalign{\smallskip}
\multicolumn{1}{c}{$\phi$} & \multicolumn{1}{c}{$\phantom{0}\psi$} &
\multicolumn{1}{c}{$\phantom{0}\psi$} &
\multicolumn{1}{c}{TN}  & 
\multicolumn{1}{c}{TK}  & 
\multicolumn{1}{c}{UN}  \\
\noalign{\smallskip}
\hline\noalign{\smallskip}                                                                                           
1.2673D$-$1 &                & $+$5.3464D$-$3 &     0 &   1 &  0 \\
1.5954D$-$1 &                &                &     0 &   0 &  0 \\
2.0085D$-$1 &                &                &     0 &   0 &  0 \\
2.5286D$-$1 &                &                &     0 &   0 &  0 \\
3.1833D$-$1 &                &                &     0 &   0 &  0 \\
4.0075D$-$1 & $-$1.4573D$-$0 & $-$4.9465D$-$1 &     2 &   1 &  0 \\
5.0451D$-$1 & $-$1.0802D$-$0 & $-$2.9362D$-$1 &     6 &   1 &  1 \\
6.3514D$-$1 & $-$4.9596D$-$1 & $-$3.9362D$-$1 &    29 &   1 &  1 \\
7.9960D$-$1 & $-$3.7170D$-$2 & $-$1.9259D$-$1 &   105 &   2 &  2 \\
1.0066D$-$0 & $-$1.8480D$-$2 & $-$2.9259D$-$1 &   138 &   1 &  3 \\
1.2673D$-$0 & $-$7.9393D$-$3 & $-$9.1564D$-$2 &   178 &   2 &  6 \\
1.5954D$-$0 & $-$2.7200D$-$1 & $-$1.9156D$-$1 &   122 &   5 &  3 \\
2.0085D$-$0 & $-$8.5630D$-$1 & $-$7.1753D$-$1 &    40 &   0 &  3 \\
2.5286D$-$0 & $-$1.9563D$-$0 & $-$1.2947D$-$0 &     4 &   0 &  1 \\
\noalign{\smallskip}
\hline                                                      
\end{tabular}                                               
\end{center}                                                
\end{table}                                                 

\subsection{Linear fit to the EDOD} \label{linf}

Both Ra07 and PM11 samples are biased towards low metallicities, [Fe/H$]
\appleq-1$
which, using the [O/H]-[Fe/H] relation inferred in the parent paper (Ramirez
et al. 2007), translates into $ \phi\appleq0.3$ for both the thin and the
thick disk.   On the other hand, thin disk stars exhibit oxygen abundance
above 0.3 within both Ra07n (Caimmi and Milanese 2009) and PM11n (Table
\ref{t:psph}) samples, while both Ra07k (Caimmi and Milanese 2009) and PM11k
(Table \ref{t:psph}) samples show a low-metallicity tail below 0.3.

It may safely be assumed the low-metallicity tail relates to a negligible mass
fraction for the thin and (to a lesser extent) the thick disk.   Accordingly,
sufficiently large samples listed in Table \ref{t:samp} (PM11n, Ra07n, Ra07k)
can be considered as representative for the thin or the thick disk, while the
remaining low sample (PM11k) is added for comparison only.

\begin{figure*}[t]  
\begin{center}      
\includegraphics[scale=0.8]{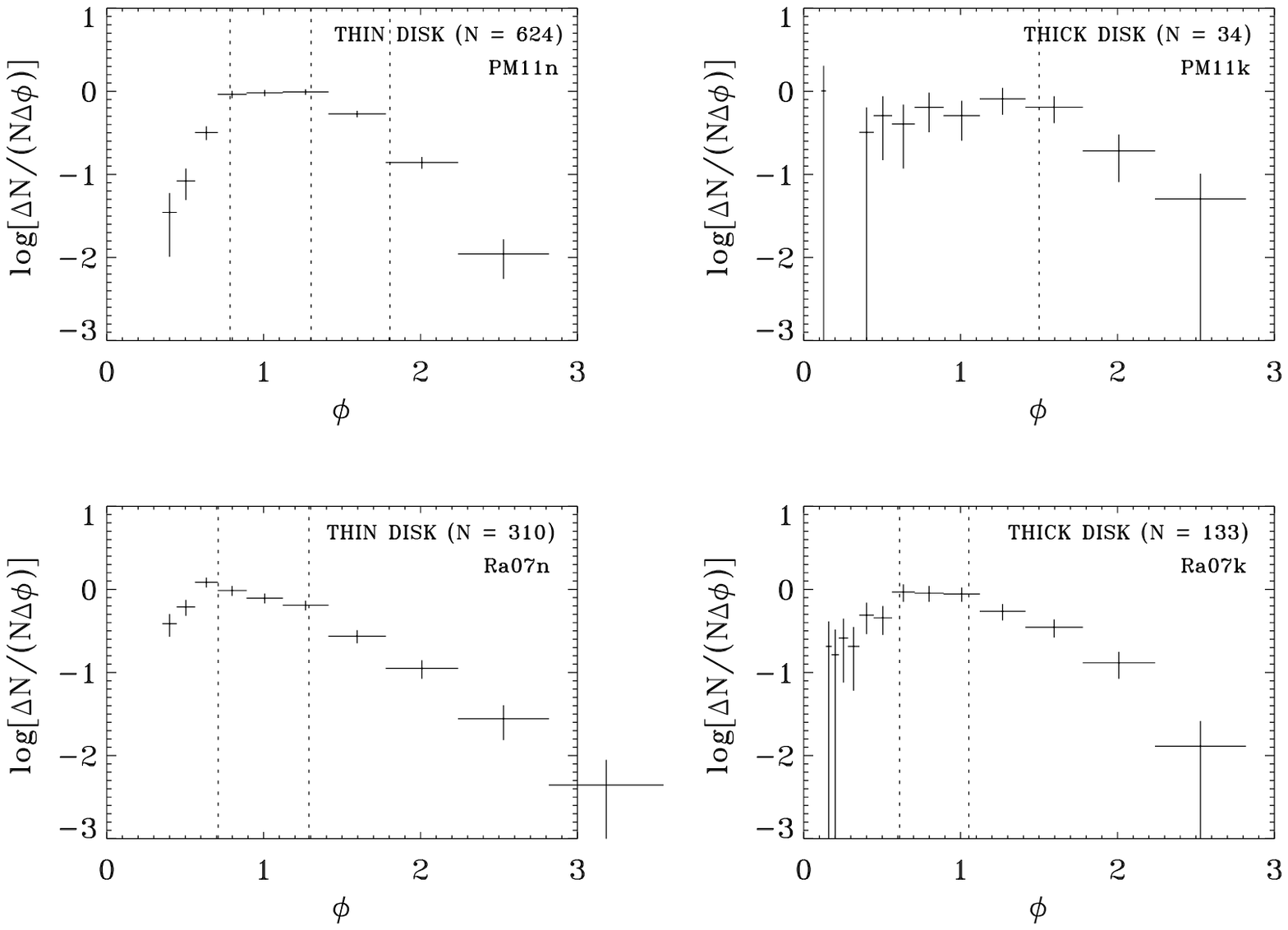}                      
\caption[ddbb]{The empirical, differential oxygen abundance
distribution (EDOD) related to PM11 (upper panels)
and Ra07 (lower panels) samples.
The uncertainty of the distribution is determined
from Poisson errors, which tend to negative infinite
for data related to a single star, including the one
outside the bottom left panel.
The vertical dotted lines mark the boundary between
adjacent regions, as determined from the intersection
of related linear fits.
For further details refer to the text.}
\label{f:pm11oh}     
\end{center}       
\end{figure*}

For all samples listed in Table \ref{t:samp}, the EDOD is plotted in
Fig.\,\ref{f:pm11oh}, where upper and lower panels represent PM11 and Ra07
samples, respectively.
Points with lower error bars attaining the horizontal axis (including the one
out of scale) correspond to bins containing a single star.

The main feature of the EDOD, plotted in Fig.\,\ref{f:pm11oh}, is the presence
of up to four regions exhibiting a nearly linear trend, which shall be named
as A, F, C, E, for increasing oxygen abundance.   Linear fits to each region,
$\psi=a\phi+b$, are performed using the bisector regression (Caimmi 2011b),
leaving aside bins related to single stars, unless the total number
of points reduces to two.   Aiming to a unified description, the number of
regions shall be kept equal to four in all cases and adjacent regions with
coinciding linear trend shall be indicated by related letters e.g., CE means
regions C and E are fitted by the same line.   The regression line slope and
intercept estimators and related dispersion estimators are listed in Table
\ref{t:abs} for each region of the EDOD plotted in Fig.\,\ref{f:pm11oh}.
\begin{table}
\caption{Regression line slope and intercept
estimators, $\hat{a}$ and $\hat{b}$, and
related dispersion estimators, $\hat{\sigma}_
{\hat{a}}$, and $\hat{\sigma}_{\hat{b}}$,
for bisector regression models applied to the 
oxygen abundance distribution (EDOD)
plotted in different panels of Fig.\,\ref{f:pm11oh}.
The method has dealt with
each region (X) separately.   Data points on the
boundary between adjacent regions are used
for determining regression lines within both
of them.}
\label{t:abs}
\begin{center}
\begin{tabular}{llllll} \hline
\multicolumn{1}{l}{X} &
\multicolumn{1}{c}{$\hat{a}$} &
\multicolumn{1}{c}{$\hat{\sigma}_{\hat{a}}$} &
\multicolumn{1}{c}{$\hat{b}$} &
\multicolumn{1}{c}{$\hat{\sigma}_{\hat{b}}$} &
\multicolumn{1}{c}{sample} \\ 
\hline
   &                 &              &                 &              &       \\
A  & $+$3.6420 E$+$0 & 1.4136 E$-$1 & $-$2.8981 E$+$0 & 6.6652 E$-$2 & PM11n \\ 
F  & $+$6.3321 E$-$2 & 5.7541 E$-$3 & $-$8.6073 E$-$2 & 5.8764 E$-$3 &       \\ 
C  & $-$1.1689 E$+$0 & 7.0570 E$-$2 & $+$1.5193 E$+$0 & 1.2997 E$-$1 &       \\ 
E  & $-$1.8294 E$+$0 & 8.0990 E$-$2 & $+$2.7116 E$+$0 & 1.8562 E$-$1 &       \\ 
   &                 &              &                 &              &       \\
A  & $+$2.1371 E$+$0 & 3.9906 E$-$2 & $-$1.2771 E$+$0 & 2.2528 E$-$2 & Ra07n \\
F  & $-$4.3676 E$-$1 & 3.0185 E$-$2 & $+$3.4846 E$-$1 & 2.8952 E$-$2 &       \\
CE & $-$1.0703 E$+$0 & 1.3474 E$-$2 & $+$1.1649 E$+$0 & 2.2887 E$-$2 &       \\
   &                 &              &                 &              &       \\
AF & $+$2.7877 E$-$1 & 7.8677 E$-$2 & $-$5.1247 E$-$1 & 7.1999 E$-$2 & PM11k \\
CE & $-$1.1799 E$+$0 & 1.8906 E$-$2 & $+$1.6773 E$+$0 & 4.1322 E$-$2 &       \\
   &                 &              &                 &              &       \\
A  & $+$1.8383 E$+$0 & 2.4630 E$-$1 & $-$1.1545 E$+$0 & 1.3187 E$-$1 & Ra07k \\
F  & $-$6.4249 E$-$2 & 2.8011 E$-$3 & $+$8.2504 E$-$3 & 2.3791 E$-$3 &       \\
CE & $-$8.1530 E$-$1 & 3.4051 E$-$2 & $+$7.8279 E$-$1 & 4.4835 E$-$2 &       \\
\hline       
\end{tabular}
\end{center} 
\end{table}  

\begin{figure*}[t]  
\begin{center}      
\includegraphics[scale=0.8]{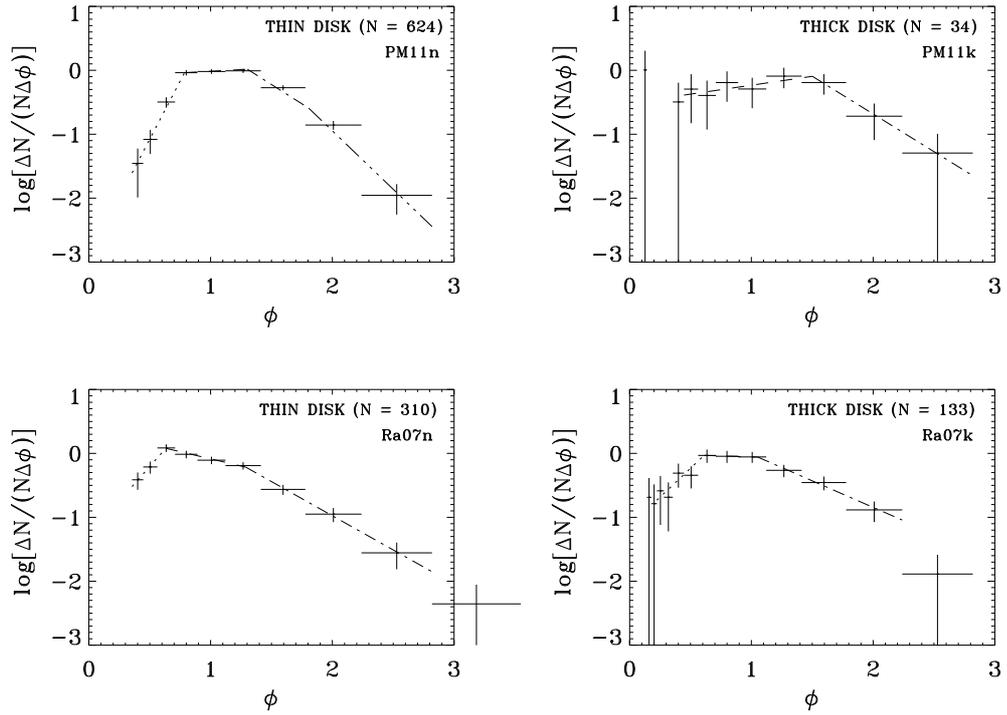}                      
\caption[ddbb]{Regression lines to the empirical
differential oxygen abundance distribution (EDOD) plotted
in Fig.\,\ref{f:pm11oh}, with regard to the regions
(from the left to the right): A, F, C, E.   If
adjacent regions exhibit a similar linear trend,
a single fit is performed.   Data related to a single
star are not used for the fitting procedure unless
only two points remain.   Other captions as in
Fig.\,\ref{f:pm11oh}.   For further details
refer to the text.}
\label{f:pm11ohl}     
\end{center}       
\end{figure*}

The regression lines are represented in Fig.\,\ref{f:pm11ohl} for each region,
according to slope and intercept values listed in Table \ref{t:abs}.   To
ensure continuity, adjacent regions are bounded by the intersections of
related regression lines.   
The results are listed in Table \ref{t:figs}, where O denotes oxygen abundance
ranges without data, $0\le\phi\le\phi_i$ and $\phi\ge\phi_f$,
where $\phi_i$ and $\phi_f$ are the minimum and maximum
oxygen abundance, respectively, within sample stars.
\begin{table}
\caption{Transition points between adjacent
regions, as determined from the intersection
of related regression lines, for the oxygen
abundance distribution (EDOD) plotted in
different panels of Fig.\,\ref{f:pm11oh}.
The repetition of the data passing from a
stage to the next one means that the interpolation
line remains unchanged and any point may be
considered as a transition point.
Cases on the left and on the right relate
to thin and thick disk, respectively (left
and right panels of Fig.\,\ref{f:pm11oh}).
Cases at the top and at the bottom relate
to PM11 and Ra07 samples, respectively (top
and bottom panels of Fig.\,\ref{f:pm11oh}).}
\label{t:figs}
\begin{center}
\begin{tabular}{lllll} \hline
\multicolumn{1}{c|}{sample}
& \multicolumn{2}{c|}{PM11n / Ra07n}
& \multicolumn{2}{c}{PM11k / Ra07k} \\
\hline
\multicolumn{1}{l|}{transition} &
\multicolumn{1}{c}{$\phi$} &
\multicolumn{1}{c|}{$\psi$} &
\multicolumn{1}{c}{$\phi$} &
\multicolumn{1}{c}{$\psi$} \\
\hline
      &              &                 &              &                 \\
O-A   & 4.0075 E$-$1 & $-$1.4386 E$+$0 & 5.0451 E$-$1 & $-$3.7183 E$-$1 \\ 
A-F   & 7.8579 E$-$1 & $-$3.6316 E$-$2 & 5.0451 E$-$1 & $-$3.7183 E$-$1 \\ 
F-C   & 1.3028 E$+$0 & $-$3.5782 E$-$3 & 1.5012 E$-$0 & $-$9.3977 E$-$2 \\ 
C-E   & 1.8052 E$+$0 & $-$5.9085 E$-$1 & 1.5012 E$-$0 & $-$9.3977 E$-$2 \\ 
E-O   & 2.5286 E$+$0 & $-$1.9141 E$-$0 & 2.5286 E$-$0 & $-$1.3061 E$-$0 \\
      &              &                 &              &                 \\
O-A   & 4.0075 E$-$1 & $-$4.2062 E$-$1 & 2.5286 E$-$1 & $-$6.8969 E$-$1 \\ 
A-F   & 6.3155 E$-$1 & $+$7.2630 E$-$2 & 6.1116 E$-$1 & $-$3.1016 E$-$2 \\ 
F-C   & 1.2887 E$-$0 & $-$2.1439 E$-$1 & 1.0313 E$-$0 & $-$5.8008 E$-$2 \\ 
C-E   & 1.2887 E$-$0 & $-$2.1439 E$-$1 & 1.0313 E$-$0 & $-$5.8008 E$-$2 \\ 
E-O   & 2.5286 E$-$0 & $-$1.5414 E$-$0 & 2.0085 E$-$0 & $-$8.5474 E$-$1 \\
\hline       
\end{tabular}
\end{center} 
\end{table}  
Accordingly, a vertical line instead of a regression line is considered for
the intersection points related to O-A and E-O transitions.

In conclusion, the EDOD related to the thin and the thick disk may be
conceived, to a satisfactory extent, as due to the contribution of four
(A, F, C, E) or three (A, F, CE) or two (AF, CE) regions within which the
trend is linear.   An interpolation in terms of MCBR models of chemical
evolution is highly attractive in that the corresponding TDOD shows, as a
special case, a linear trend (C11).

\section{The model}\label{mod}
\subsection{General remarks}\label{gerT}

Simple MCBR models have been presented in an earlier attempt (C11) and an
interested reader is addressed therein for an exhaustive formulation.   Only
what is relevant for the application shall be repeated here and related
extension and improvement shall be performed.   The main assumptions of the
model are listed below.
\begin{description}
\item[$\bullet$\hspace{6.1mm}]
System structured as a box and a reservoir where gas can be exchanged between
the two but mass conservation holds.
\item[$\bullet$\hspace{6.1mm}]
Instantaneous recycling within the box, where stars are
divided into two categories, namely
(a) short-lived, which instantaneously
evolve, and (b) long-lived, all of
which are still evolving.
\item[$\bullet$\hspace{6.1mm}]
Instantaneous mixing within the box.
\item[$\bullet$\hspace{6.1mm}]
Gas outflow from the box into the reservoir
or inflow into the box from the reservoir,
at a rate proportional to the star formation
rate and with fixed composition.
\item[$\bullet$\hspace{6.1mm}]
Inhibition of star formation within the reservoir.
\item[$\bullet$\hspace{6.1mm}]
Oxygen abundance proportional to metal abundance, according to the forthcoming
Eq.\,(\ref{eq:AO}).
\end{description}

In this picture, the gas mass fraction, $\mu$, (normalized to the initial
mass) and the oxygen abundance, $\phi$, (normalized to the solar value) are
related as:
\begin{lefteqnarray}
\label{eq:mug}
&& \frac{\mu}{\mu_i}=\left[\frac{1-c\phi}{1-c\phi_i}\right]^{(Z_{\rm O})_
\odot/(c\hat{p}^\pprime)}~~; \\
\label{eq:c}
&& c=\frac{(Z_{\rm O})_\odot}{\hat{p}}\left[A_{\rm O}\hat{p}-\kappa(1-\zeta_
{\rm O})\right]~~; \\
\label{eq:ps}
&& \hat{p}^\pprime=\frac{\hat{p}}{1+\kappa}~~; \\
\label{eq:AO}
&& Z=A_{\rm O}Z_{\rm O}~~;
\end{lefteqnarray}
where $(Z_{\rm O})_\odot$ is the solar oxygen abundance, 
$\hat{p}$ is the yield, $\hat{p}^\pprime$ the effective
yield, $\kappa$ the flow (positive for outflow and negative for inflow),
$\zeta_{\rm O}$ the cut, $Z_{\rm O}$ the oxygen abundance,  $Z$ the metal
abundance, $A_{\rm O}=2$ to an acceptable extent and the index, $i$, denotes
values at the beginning of evolution.

More specifically, $\kappa$ represents the ratio of gas mass locked into
long-lived stars and stellar remnants to gas mass outflowed into the reservoir
or inflowed from the reservoir and $\zeta_{\rm O}$ represents the ratio of
oxygen abundance within the flowing gas to oxygen abundance within the
pre existing gas.   For further details refer to the parent paper (C11).
In principle, $-\infty<\kappa<+\infty$ and $0\le\zeta_{\rm O}\le A_{\rm O}/
(Z_{\rm O})_f$.   
The domain, $\zeta_{\rm O}>A_{\rm O}/(Z_{\rm O})_f$, is in contradiction
with the assumption of fixed $\zeta_{\rm O}$ during the evolution.   

The TDOD predicted by the model reads:
\begin{lefteqnarray}
\label{eq:psit}
&& \psi=\left[\frac{(Z_{\rm O})_\odot}{c\hat{p}^\pprime}-1\right]
\log(1-c\phi)+b~~; \\
\label{eq:b}
&& b=\log\left[\frac{\mu_i}{\mu_i-\mu_f}\frac{(Z_{\rm O})_\odot}{\hat{p}^
\pprime}\right]-\frac{(Z_{\rm O})_\odot}{c\hat{p}^\pprime}\log(1-c\phi_i)~~;
\end{lefteqnarray}
where $b$ is the intercept of the curve on the $({\sf O}\phi\psi)$ plane.
For further details refer to the parent paper (C11).

\subsection{Theoretical integral oxygen abundance distribution (TIOD)}
\label{TIOD}

An explicit expression of the TIOD could be needed for a number of
applications.
The theoretical counterpart of the EDOD, via Eq.\,(\ref{eq:psie}), reads:
\begin{lefteqnarray}
\label{eq:pstt}
&& \psi=\log\frac{\diff N}{(N_f-N_i)\diff\phi}~~;
\end{lefteqnarray}
where $N_i$, $N_f$, denote the total number of long-lived stars with
normalized oxygen abundance, $\phi\le\phi_i$, $\phi\le\phi_f$, respectively.

The combination of Eqs.\,(\ref{eq:psit}) and (\ref{eq:pstt}) yields after some
algebra:
\begin{lefteqnarray}
\label{eq:ptdf}
&& \frac{\diff N}{N_f-N_i}=10^b(1-c\phi)^{(Z_{\rm O})_\odot/
(c\hat{p}^\pprime)-1}\diff\phi~~;
\end{lefteqnarray}
which can be integrated.   The result is:
\begin{lefteqnarray}
\label{eq:ptin}
&& \frac{N-N_i}{N_f-N_i}=-10^b\frac{\hat{p}^\pprime}{(Z_{\rm O})_\odot}\left[
(1-c\phi)^{(Z_{\rm O})_\odot/(c\hat{p}^\pprime)}-(1-c\phi_i)^{(Z_{\rm O})_
\odot/(c\hat{p}^\pprime)}\right]~~;
\end{lefteqnarray}
where $N$ is the total number of stars with normalized oxygen abundance not
exceeding $\phi$.   Accordingly, the TIOD is expressed by
Eq.\,(\ref{eq:ptin}).

\subsection{A family of TDOD curves}\label{fatc}

The TDOD, expressed by Eq.\,(\ref{eq:psit}), via (\ref{eq:mug}), (\ref{eq:c}),
(\ref{eq:ps}), (\ref{eq:b}), depends on the input parameters, $[A_{\rm O},
\hat{p}, (Z_{\rm O})_\odot, \phi_i, \phi_f, \mu_i]$ and the output parameters,
$(\kappa, \zeta_{\rm O})$.   With regard to a selected stage of evolution (or
adjacent stages fitted by the same line), let $(\phi_i,\psi_i)$ and $(\phi_f,
\psi_f)$ be the starting and the ending point, respectively, of the related
linear fit.
The family of TDOD curves, passing through the above mentioned points,
necessarily satisfy the relations:
\begin{lefteqnarray}
\label{eq:psii}
&& \psi_i=\left[\frac{(Z_{\rm O})_\odot}{c\hat{p}^\pprime}-1\right]
\log(1-c\phi_i)+b~~; \\
\label{eq:psif}
&& \psi_f=\left[\frac{(Z_{\rm O})_\odot}{c\hat{p}^\pprime}-1\right]
\log(1-c\phi_f)+b~~;
\end{lefteqnarray}
which makes a system of two equations in the two unknowns, $(\kappa,
\zeta_{\rm O})$.

The combination of Eqs.\,(\ref{eq:psii}) and (\ref{eq:psif}) yields after some
algebra:
\begin{lefteqnarray}
\label{eq:chint}
&& \kappa=\frac{c\hat{p}}{(Z_{\rm O})_\odot}\left[1+(\psi_f-\psi_i)\left(\log
\frac{1-c\phi_f}{1-c\phi_i}\right)^{-1}\right]-1~~;
\end{lefteqnarray}
which, for assigned $\zeta_{\rm O}$, can be solved by repeated iterations
using Eq.\,(\ref{eq:c}) until $\vert\kappa^{(n)}-\kappa^{(n-1)}\vert<
\epsilon$, where $n$ is the last iteration and $\epsilon$ a fixed threshold.
Then the generic curve of the family is defined by the parameters, $(\kappa,
\zeta_{\rm O})$.   The intercept, $b$, is determined via Eq.\,(\ref{eq:psii})
or (\ref{eq:psif}).

Both the EDOD and the TDOD can be represented on the $({\sf O}\phi\psi)$
plane, the former and the latter normalized to the long-lived star population
of the sample and the environment under consideration, respectively, which
translates into a normalization constant, $\log C_{\rm N}$.   More
specifically, the TDOD has to be vertically shifted on the $({\sf O}\phi\psi)$
plane by a value, $\log C_{\rm N}$, for matching to the EDOD.   The result is:
\begin{equation}
\label{eq:CN}
C_{\rm N}=10^b\frac{\hat{p}^\pprime}{(Z_{\rm O})_\odot}\frac{\mu_i-\mu_f}
{\mu_i}(1-c\phi_i)^{(Z_{\rm O})_\odot/(c\hat{p}^\pprime)}~~;
\end{equation}
where $b$ is defined by Eq.\,(\ref{eq:b}).   For further details refer to the
parent paper (C11), keeping in mind the family of TDOD curves defined therein
have in common the points, $(0,b)$ and $(\phi_f,\psi_f)$,  instead of
$(\phi_i,\psi_i)$ and $(\phi_f,\psi_f)$, respectively.

\subsection{The linear limit}\label{lili}

In the special case, $c=0$, Eq.\,(\ref{eq:c}) reduces to:
\begin{equation}
\label{eq:zilt}
\zeta_{\rm O}=1-\frac{A_{\rm O}\hat{p}}\kappa~~;
\end{equation}
which implies $\zeta_{\rm O}>1$ for $\kappa<0$; $0\le\zeta_{\rm O}<1$ for
$\kappa\ge A_{\rm O}\hat{p}$; $\zeta_{\rm O}=1$ for $\vert\kappa\vert\to+
\infty$; $\zeta_{\rm O}<0$ for $0\le\kappa<A_{\rm O}\hat{p}$.   The last
subdomain has to be excluded as negative cut values, by definition, have no
physical meaning.

With regard to the remaining domain, Eqs.\,(\ref{eq:mug}), (\ref{eq:psit}),
(\ref{eq:b}), (\ref{eq:ptin}), reduce to:
\begin{lefteqnarray}
\label{eq:fimfl}
&& \frac\mu{\mu_i}=\exp\left[-\frac{(Z_{\rm O})_\odot}{\hat{p}^\pprime}
(\phi-\phi_i)\right]~~; \\
\label{eq:psilt}
&& \psi=a\phi+b~~; \\
\label{eq:bl}
&& b=\log\left[\frac{\mu_i}{\mu_i-\mu_f}(-\ln10)a\right]-a\phi_i~~; \\
\label{eq:al}
&& a=-\frac1{\ln10}\frac{(Z_{\rm O})_\odot}{\hat{p}^\pprime}~~; \\
\label{eq:ptil}
&& \frac{N-N_i}{N_f-N_i}=\frac1{\ln10}\frac1a\left[\exp_{10}(a\phi+b)-
\exp_{10}(a\phi_i+b)\right]~~;
\end{lefteqnarray}
implying a linear TDOD on the $({\sf O}\phi\psi)$ plane.   Accordingly, the
limit, $c\to0$, shall be quoted as the linear limit.

Keeping in mind the starting and the ending point of the linear fit are
$(\phi_i,\psi_i)$, $(\phi_f,\psi_f)$, respectively, an empirical counterpart
of Eq.\,(\ref{eq:psilt}) reads:
\begin{equation}
\label{eq:psile}
\psi=\psi_i+\frac{\psi_f-\psi_i}{\phi_f-\phi_i}(\phi-\phi_i)~~;
\end{equation}
which implies, by comparison with Eq.\,(\ref{eq:psilt}) via (\ref{eq:ps}) and
(\ref{eq:al}):
\begin{lefteqnarray}
\label{eq:af}
&& a=\frac{\psi_f-\psi_i}{\phi_f-\phi_i}=a_0(1+\kappa)~~; \\
\label{eq:a0}
&& a_0=-\frac1{\ln10}\frac{(Z_{\rm O})_\odot}{\hat{p}}~~;
\end{lefteqnarray}
where $a_0$ is the slope related to stagnation regime, $\kappa=0$.
Accordingly, $\kappa=a/a_0-1$ and Eq.\,(\ref{eq:zilt}) takes the expression:
\begin{equation}
\label{eq:zil}
\zeta_{\rm O}=1+\frac{A_{\rm O}\hat{p}}{1-a/a_0}~~;
\end{equation}
where the output parameters, $(\kappa,\zeta_{\rm O})$, may be
determined using Eqs.\,(\ref{eq:af})-(\ref{eq:zil}).
%
%
%

\subsection{The dominant flow limit}\label{dofl}

In the limit of negligible star formation rate with respect to flow rate,
$\vert\kappa\vert\to+\infty$, the following relations hold after some algebra:
\begin{lefteqnarray}
\label{eq:li1}
&& \lim_{\vert\kappa\vert\to+\infty}\left[\frac{(Z_{\rm O})_\odot}{c\hat{p}^
\pprime}\right]=-\frac1{1-\zeta_{\rm O}}~~; \\
\label{eq:li2}
&& \lim_{\vert\kappa\vert\to+\infty}\left[\frac{(Z_{\rm O})_\odot}{\hat{p}^
\pprime}\frac1{1-c\phi_i}\right]=\frac1{1-\zeta_{\rm O}}\frac1{\phi_i}~~; \\
\label{eq:li3}
&& \lim_{\vert\kappa\vert\to+\infty}\left[\log\frac{1-c\phi}{1-c\phi_i}\right]=
\log\frac\phi{\phi_i}~~; \\
\label{eq:li4}
&& \lim_{\vert\kappa\vert\to+\infty}\left\{b+\left[\frac{(Z_{\rm O})_\odot}
{c\hat{p}^\pprime}-1\right]\log(1-c\phi_i)\right\}=\log\left[\frac{\mu_i}
{\mu_i-\mu_f}\frac1{1-\zeta_{\rm O}}\frac1{\phi_i}\right]~~;\qquad
\end{lefteqnarray}
and Eqs.\,(\ref{eq:mug}) and (\ref{eq:psit}) reduce to:
\begin{lefteqnarray}
\label{eq:li5}
&& \lim_{\vert\kappa\vert\to+\infty}\frac\mu{\mu_i}=\left(\frac\phi{\phi_i}
\right)^{-1/(1-\zeta_{\rm O})}~~; \\
\label{eq:li6}
&& \lim_{\vert\kappa\vert\to+\infty}\psi=\frac{\zeta_{\rm O}-2}
{1-\zeta_{\rm O}}\log\frac\phi{\phi_i}+\log\left[\frac{\mu_i}{\mu_i-\mu_f}
\frac1{1-\zeta_{\rm O}}\frac1{\phi_i}\right]~~;
\end{lefteqnarray}
where the last relation implies a divergence for the intercept,
$\psi(0)\to\infty$.

The particularization of Eq.\,(\ref{eq:li6}) to the points, $(\phi_i,\psi_i)$,
$(\phi_f,\psi_f)$, yields:
\begin{equation}
\label{eq:dps}
\psi_f-\psi_i=\frac{\zeta_{\rm O}-2}{1-\zeta_{\rm O}}\log\frac{\phi_f}{\phi_i}
~~;
\end{equation}
accordingly, the cut, $\zeta_{\rm O}$, reads after some algebra:
\begin{lefteqnarray}
\label{eq:zis}
&& \zeta_{\rm O}=\frac{(\psi_f-\psi_i)+2\log(\phi_f/\phi_i)}{(\psi_f-\psi_i)+
\log(\phi_f/\phi_i)}=\frac{a/\gamma+2}{a/\gamma+1}~~; \\
\label{eq:cin}
&& \gamma=\frac{\log(\phi_f/\phi_i)}{\phi_f-\phi_i}~~;
\end{lefteqnarray}
where $a$ is the slope of the linear fit, expressed by Eq.\,(\ref{eq:af}), and
$\gamma>0$.


The intercept, $b$, according to Eq.\,(\ref{eq:li4}), can be expressed as:
\begin{lefteqnarray}
\label{eq:bbp}
&& b=b^\prime-\left[\frac{(Z_{\rm O})_\odot}{c\hat{p}^\pprime}-1\right]\log(1-c
\phi_i)~~; \\
\label{eq:bp}
&& b^\prime=\log\left[\frac{\mu_i}{\mu_i-\mu_f}\frac1{1-\zeta_{\rm O}}\frac1
{\phi_i}\right]~~;
\end{lefteqnarray}
where $\kappa\to\mp\infty$ necessarily implies $\mu_i\legr\mu_f$ and, in turn,
$\zeta_{\rm O}\grle1$, within the domain of the reduced intercept, $b^\prime$,
expressed by Eq.\,(\ref{eq:bp}). 
Keeping in mind Eq.\,(\ref{eq:c}), the last term on the right-hand side of
Eq.\,(\ref{eq:bbp}) in the case under discussion shows a divergence as:
\begin{equation}
\label{eq:lib}
\lim_{\vert\kappa\vert\to+\infty}\left\{\left[\frac{(Z_{\rm O})_\odot}
{c\hat{p}^\pprime}-1\right]\log(1-c\phi_i)\right\}=\frac{\zeta_{\rm O}-2}
{1-\zeta_{\rm O}}\lim_{\vert\kappa\vert\to+\infty}\log(1-c\phi_i)=\infty~~;
\end{equation}
accordingly, the normalization constant, $C_{\rm N}$, must be related to the
reduced intercept, $b^\prime$, whose empirical counterpart equals $\psi_i$ via
Eqs.\,(\ref{eq:psii}) and (\ref{eq:bbp}).   Then Eq.\,(\ref{eq:CN}) translates
into:
\begin{equation}
\label{eq:CNin}
C_{\rm N}=10^{\psi_i}\frac{\mu_i-\mu_f}{\mu_i}(1-\zeta_{\rm O})\phi_i~~;
\end{equation}
which is equivalent to $\psi_i=b^\prime+\log C_{\rm N}$.

\section{The $({\sf O}\kappa\zeta_{\rm O})$ plane}
\label{OkzO}

\begin{figure*}[t]  
\begin{center}      
\includegraphics[scale=0.8]{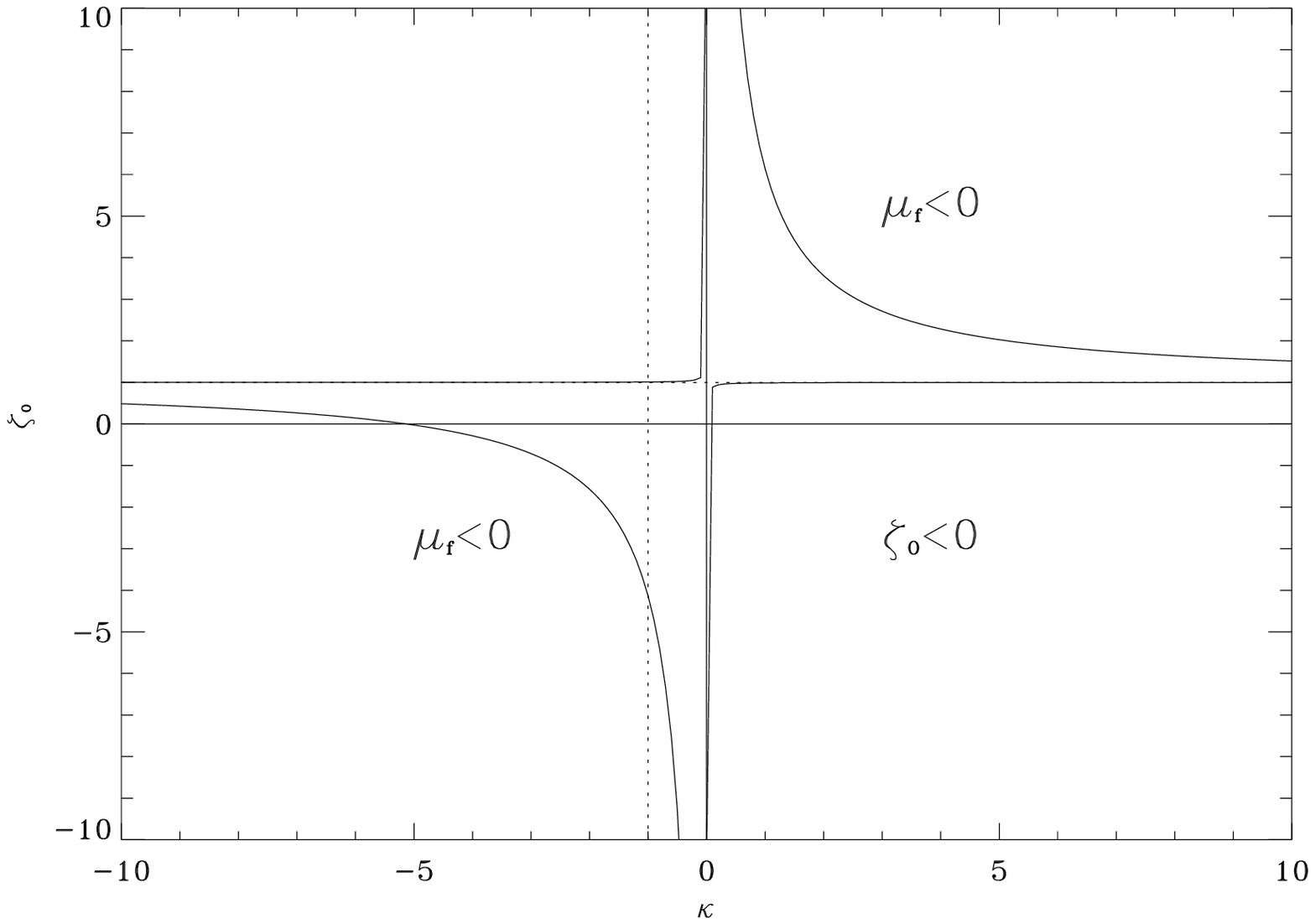}                      
\caption[ddbb]{The  parameter space on the $({\sf O}\kappa\zeta_{\rm O})$
plane.   The locus,
$c=0$, is represended by the equilateral hyperbola lying on the second and
fourth quadrant, where $c>0$ or $c<0$ for points placed between the branches
of the hyperbola or outside either branch, respectively.   The locus,
$\mu_f=0$, is represended by the equilateral hyperbola lying on the first and
third quadrant, where $\mu_f>0$ or $\mu_f<0$ for points placed between the
branches of the hyperbola or outside either branch, respectively.   The
vertical axis, $\kappa=0$, and the horizontal axis, $\zeta_{\rm O}=1$, are
asymptotes for both hyperbolas.   The vertical axis, $\kappa=-1$, represents
the steady inflow regime.   The parameter space is restricted to
the region bounded by the branches of the
hyperbola, $\mu_f=0$, and the horizontal axis, $\zeta_{\rm O}=0$, which
ensures $\mu_f\ge0$ and $\zeta_{\rm O}\ge0$.
For further details refer to the text.}
\label{f:hype}     
\end{center}       
\end{figure*}

The linear limit, $c=0$, is represented on the $({\sf O}\kappa\zeta_{\rm O})$
plane, via Eq.\,(\ref{eq:zilt}), as an equilateral hyperbola of asymptotes,
$\kappa=0$ and $\zeta_{\rm O}=1$, shown in Fig.\,\ref{f:hype}.
It can be seen $c>0$ for points lying between the branches of the hyperbola,
while $c<0$ for points lying outside either branch of the hyperbola.

The steady inflow regime, $\kappa=-1$, is represented on the $({\sf O}\kappa
\zeta_{\rm O})$ plane as a vertical axis.   It can be seen $\mu_f>\mu_i$ in
the strong inflow regime, $\kappa<-1$, and $\mu_f<\mu_i$ in the weak inflow
regime, $-1<\kappa<0$, or in the stagnation regime, $\kappa=0$, or in outflow
regime, $\kappa>0$, while $\mu_f=\mu_i$ in the steady inflow regime.

The limit of complete gas exhaustion, via Eq.\,(\ref{eq:mug}), reads $c=1/\phi
_f$ which, using Eq.\,(\ref{eq:c}), can be cast under the explicit form:
\begin{equation}
\label{eq:muf0}
\zeta_{\rm O}=1-\left[A_{\rm O}\hat{p}-\frac{\hat{p}}{(Z_{\rm O})_\odot}\frac1
{\phi_f}\right]\frac1\kappa~~;
\end{equation}
that is an equilateral hyperbola with asymptotes, $\kappa=0$, $\zeta_{\rm O}=1
$, as shown in Fig.\,\ref{f:hype}.
It can be seen $\mu_f>0$ for points lying between the branches of the
hyperbola, while $\mu_f<0$ for points lying outside either branch of the
hyperbola.   The latter implies lack of physical meaning, together with points
lying on the negative semiplane, $\zeta_{\rm O}<0$.    Accordingly, the
parameter space on the $({\sf O}\kappa\zeta_{\rm O})$ plane is restricted to
the region bounded by the branches of the hyperbola, expressed by
Eq.\,(\ref{eq:muf0}), and the horizontal axis, $\zeta_{\rm O}=0$,
where $\mu_f\ge0$, $\zeta_{\rm O}\ge0$, as shown in Fig.\,\ref{f:hype}.

It is worth noticing $c\phi_f\to1$ implies $\vert\psi(\phi_f)\vert\to+\infty$
via Eq.\,(\ref{eq:psit}) and the related curve cannot belong to the family,
defined by Eqs.\,(\ref{eq:psii}) and (\ref{eq:psif}).

\section{Results}
\label{resu}

The EDOD related to the samples considered in Section 2 can be divided into
four, three, or two regions, where the trend is linear to a good extent, as
shown in Fig.\,\ref{f:pm11ohl}.   In the light of the model, each region
corresponds to a different stage of chemical evolution, which can tentatively
be related to a different stage of dynamical evolution.   For further details
refer to the parent paper (C11).

The fractional mass of the box (with respect to the initial value) attains the
maximum value at the end of the last stage, P, where the TDOD still shows a
slope, $a>a_0$, i.e. inflow regime (into the box from the reservoir).   The
related value is $(\mu_{\rm P})_f+(s_{\rm P})_f$.   The fractional stellar
mass of the box at the end of evolution, which coincides with the end of the
last stage, L, is $(s_{\rm L})_f$.   Accordingly, the mass ratio of the box at
the end of evolution to the outflowed gas reads:
\begin{equation}
\label{eq:Mbo}
\frac{M_{\rm box}}{M_{\rm ofl}}=\frac{(\mu_{\rm L})_f+(s_{\rm L})_f}
{(\mu_{\rm P})_f+(s_{\rm P})_f-(\mu_{\rm L})_f-(s_{\rm L})_f}~~;
\end{equation}
provided earlier stages with respect to P are in inflow regime and later
stages in outflow regime.

Following the procedure outlined in the parent paper (C11) and keeping the
same values of the input parameters, $(Z_{\rm O})_\odot=0.0056$, $\hat{p}/
(Z_{\rm O})_\odot=1.0340$, $\mu_i=1$, $s_i=0$; $D_i=0$, the normalization
constant, the cut, the flow, the active gas mass fraction, the star mass
fraction and the inflowed or outflowed gas mass fraction, at the end of each
stage of evolution, can be computed.   The results are listed in Tables
\ref{t:zita} and \ref{t:ralm}, where different cases are
related to different EDOD data plotted in Fig.\,\ref{f:pm11oh}.   The index,
U, denotes a stage (or adjacent stages) of evolution where the trend is linear
to a good extent and the indexes, lin and $\infty$, mark the linear and the
dominant flow limit, respectively.
\begin{table}
\caption{Output parameters, $[(\zeta_{\rm O})_{\rm U}]_{\rm lin}$ and
$[(\zeta_{\rm O})_{\rm U}]_\infty$,
related to the linear limit (lin) 
and to the dominant flow limit $(\infty)$, for
simple MCBR models where the theoretical differential oxygen abundance
distribution (TDOD) provides a linear fit to the empirical differential
oxygen abundance distribution (EDOD)
plotted in different panels of Fig.\,\ref{f:pm11oh}.
Four, three, or two
stages of evolution are
considered, according to the linear
trends exhibited by the EDOD.}
\label{t:zita}
\begin{center}
\begin{tabular}{llll} \hline
\multicolumn{1}{l}{U} &
\multicolumn{1}{c}{$[(\zeta_{\rm O})_{\rm U}]_{\rm lin}$} &
\multicolumn{1}{c}{$[(\zeta_{\rm O})_{\rm U}]_\infty$} &
\multicolumn{1}{c}{sample} \\
\hline

      &             &             &       \\
A     & 1.0012E$+$0 & 1.1726E$+$0 & PM11n \\
F     & 1.0101E$+$0 & 1.8702E$+$0 &       \\
C     & 9.9351E$-$1 & 6.8213E$-$1 &       \\
E     & 9.9655E$-$1 & 8.7566E$-$1 &       \\
      &             &             &       \\
A     & 1.0019E$+$0 & 1.2860E$+$0 & Ra07n \\
F     & 7.0971E$-$1 & 1.4628E$+$1 &       \\
CE    & 9.9252E$-$1 & 7.1698E$-$1 &       \\
      &             &             &       \\
AF    & 1.0070E$+$0 & 1.6302E$+$0 & PM11k \\
CE    & 9.9360E$-$1 & 7.7029E$-$1 &       \\
      &             &             &       \\
A     & 1.0023E$+$0 & 1.3778E$+$0 & Ra07k \\
F     & 1.0138E$+$0 & 2.1469E$+$0 &       \\
CE    & 9.8770E$-$1 & 4.2840E$-$1 &       \\
\hline                            
\end{tabular}                     
\end{center}                      
\end{table}                       
\begin{table}
\caption{Output parameters, $(C_{\rm U})_{\rm N}$,
$\kappa_{\rm U}$, $(\mu_{\rm U})_f$,
$(s_{\rm U})_f$, $(D_{\rm U})_f$, for
simple MCBR models where the theoretical differential oxygen abundance
distribution (TDOD) provides a linear fit to the empirical differential
oxygen abundance distribution (EDOD)
plotted in different panels of Fig.\,\ref{f:pm11oh}.
Other captions as in Table \ref{t:zita}.}
\label{t:ralm}
\begin{center}
\begin{tabular}{llllll} \hline
\multicolumn{1}{l}{U} &
\multicolumn{1}{c}{$(C_{\rm U})_{\rm N}$} &
\multicolumn{1}{c}{$\kappa_{\rm U}$} &
\multicolumn{1}{c}{$(\mu_{\rm U})_f$} &
\multicolumn{1}{c}{$(s_{\rm U})_f$} &
\multicolumn{1}{c}{$(D_{\rm U})_f$} \\
\hline

      &             &                &             &             &                \\
A     & 1.0534E$-$1 & $-$9.6714E$+$0 & 2.5253E$+$1 & 2.7969E$+$0 & $-$2.7050E$+$1 \\
F     & 4.9391E$-$1 & $-$1.1508E$+$0 & 2.7230E$+$1 & 1.5911E$+$1 & $-$4.2141E$+$1 \\
C     & 2.7317E$-$1 & $+$1.7831E$+$0 & 7.0432E$+$0 & 2.3164E$+$1 & $-$2.9207E$+$1 \\
E     & 5.8008E$-$2 & $+$3.3557E$+$0 & 3.3456E$-$1 & 2.4704E$+$1 & $-$2.4039E$+$1 \\
      &             &                &             &             &                \\
A     & 1.6306E$-$1 & $-$6.0883E$+$0 & 3.1135E$+$0 & 4.1536E$-$1 & $-$2.5289E$+$0 \\
F     & 5.6841E$-$1 & $+$3.9895E$-$2 & 1.6078E$+$0 & 1.8633E$+$0 & $-$2.4711E$+$0 \\
CE    & 2.3602E$-$1 & $+$1.5483E$+$0 & 7.5727E$-$2 & 2.4645E$+$0 & $-$1.5402E$+$0 \\
      &             &                &             &             &                \\
AF    & 5.9299E$-$1 & $-$1.6637E$+$0 & 1.8961E$+$0 & 1.3500E$+$0 & $-$2.2461E$+$0 \\
CE    & 8.7126E$-$1 & $+$1.8093E$+$0 & 1.1633E$-$1 & 1.9835E$+$0 & $-$1.0999E$+$0 \\
      &             &                &             &             &                \\
A     & 1.7169E$-$1 & $-$5.3769E$+$0 & 4.5570E$+$0 & 8.1266E$-$1 & $-$4.3696E$+$0 \\
F     & 5.5095E$-$1 & $-$8.4703E$-$1 & 4.2824E$+$0 & 2.6077E$+$0 & $-$5.8901E$+$0 \\
CE    & 9.4260E$-$1 & $+$9.4119E$-$1 & 6.8384E$-$1 & 4.4615E$+$0 & $-$4.1454E$+$0 \\
\hline                            
\end{tabular}                     
\end{center}                      
\end{table}                       

The mass ratio of the box at the end of evolution to the outflowed gas,
expressed by Eq.\,(\ref{eq:Mbo}), and the box mass fraction at the end of
evolution, $(M_{\rm box})_f/(M_{\rm box})_i=(\mu_{\rm L})_f+(s_{\rm L})_f$,
can be inferred from Table \ref{t:ralm}, where P = F, A, AF, and L = E, CE,
according to the case.   The result is (leaving aside the
poorly populated PM11k sample) $M_{\rm box}/M_{\rm ofl}=$1.4-2.6 for the
thin disk and 2.9 for the thick disk; $(M_{\rm box})_f/(M_{\rm box})_i
=$2.5-25.0 for the thin disk and 5.6 for the thick disk.

The validity of the linear limit implies cut values very close to unity,
$\zeta_{\rm O}\approx1$, unless $\kappa\to0^-$ or $\kappa\to(A_{\rm O}\hat{p})
^+$ according to Eq.\,(\ref{eq:zilt}), as can be seen in Table \ref{t:ralm}.
On the other hand, inflowing gas from intergalactic medium is expected to be
oxygen-poor with respect to galaxies, while outflowing gas into galactic
medium is expected to be slightly oxygen-rich with respect to pre existing
gas.   For this reason, the departure from the linear limit must also be
considered.

\begin{figure*}[t]  
\begin{center}      
\includegraphics[scale=0.8]{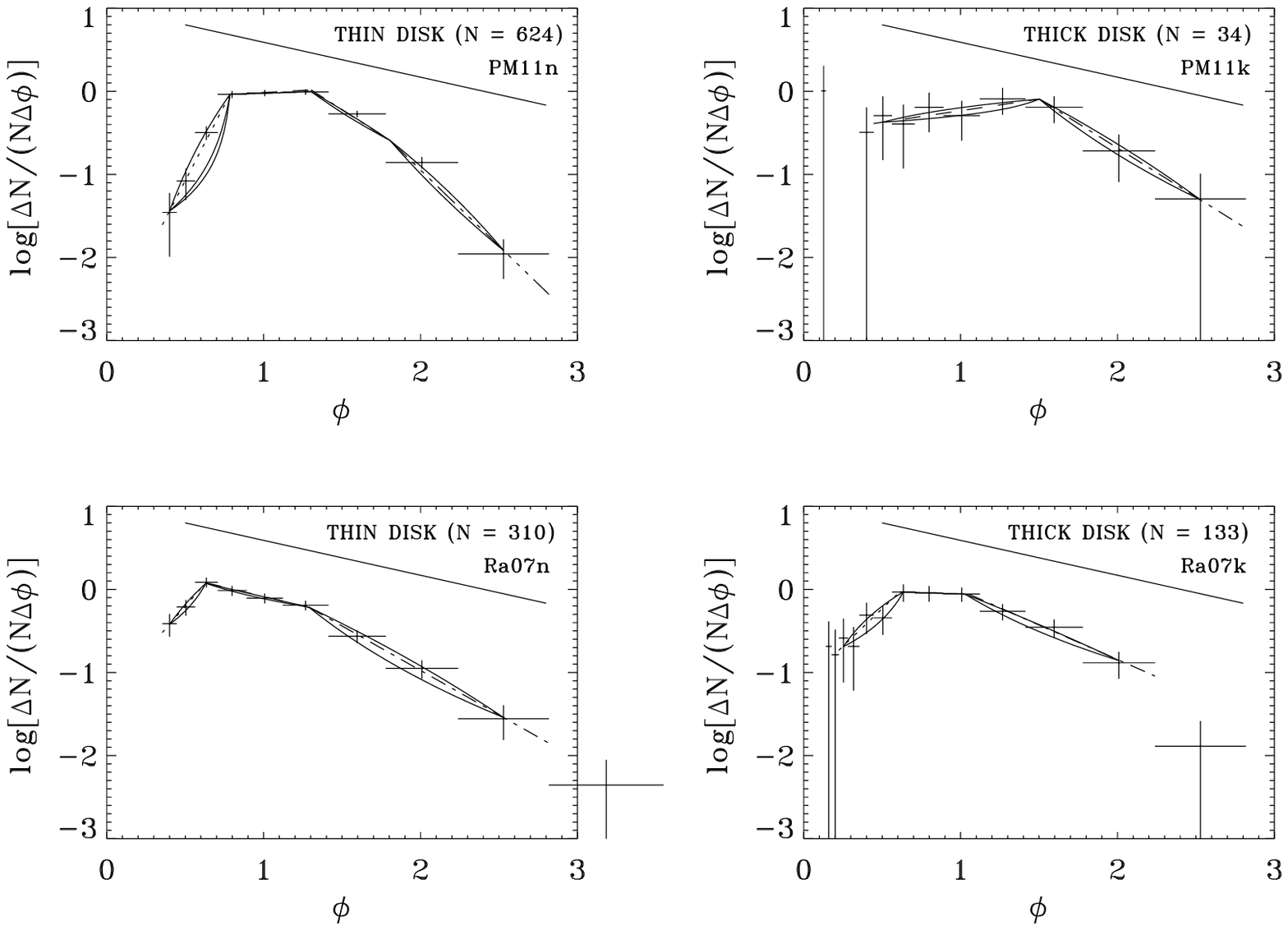}                      
\caption[ddbb]{Comparison between empirical (EDOD), plotted in
Fig.\,\ref{f:pm11oh}, and
theoretical (TDOD) differential oxygen abundance
distribution defined by the family of curves including
the regression line and passing through the starting point,
$(\phi_i,\psi_i)$, and the ending point, $(\phi_f,\psi_f)$,
where $\psi_u=a\phi_u+b_{\rm N}$, $u=i,f,$ $b_{\rm N}=b+\log C_{\rm N}$, for
the stage under consideration.   Curves correspond
(from bottom to top) to cut values, $\zeta_{\rm O}$, between 0
and $(\zeta_{\rm O})_\infty$ or $(\zeta_{\rm O})_\infty$ and 1.2
for negative and positive flow values, $\kappa$, respectively.
With regard to the top left panel, the lower curve fitting the
data relates to $\zeta_{\rm O}=0.35$.
In general, for varying cut,
$(\zeta_{\rm O})_{\rm min}\le\zeta_{\rm O}\le(\zeta_{\rm O})_{\rm max}$,
the TDOD lies within the region bounded by the curves,
$\zeta_{\rm O}=(\zeta_{\rm O})_{\rm min}$ and $\zeta_
{\rm O}=(\zeta_{\rm O})_{\rm max}$.   The full straight line on the top
of each panel has slope, $a_0=-(1/\ln10)[(Z_{\rm O})_\odot/\hat{p}]=-0.42$,
and relates to null flow, $\kappa=0$, thus separating the inflow regime,
$a>a_0$, $\kappa<0$, from the outflow regime, $a<a_0$, $\kappa>0$.  A
horizontal straight line corresponds to the steady inflow regime, $a=0$,
$\kappa=-1$.
}
\label{f:pm11oht}     
\end{center}       
\end{figure*}

The results are shown in Fig.\,\ref{f:pm11oht} where, in addition to the
linear limit shown in Fig.\,\ref{f:pm11ohl}, the following cases are also
represented: $\zeta_{\rm O}=0, (\zeta_{\rm O})_\infty$, for $\kappa<0$ i.e.
inflow regime and $\zeta_{\rm O}=(\zeta_{\rm O})_\infty$, 1.2, for $\kappa>0$
i.e. outflow regime.   In general, curves from up to down relate to decreasing
cut, $\zeta_{\rm O}$.   All curves maintain close to the related linear limit
and fit to the data, with the exception of the A stage corresponding to the
PM11n sample, where curves within the range, $0\le\zeta_{\rm O}<0.35$, do not
match the whole set of error boxes.   The curve related to $\zeta_{\rm O}=
0.35$ is also plotted on the top left panel of Fig.\,\ref{f:pm11oht}.

With regard to the PM11n sample restricted to the A stage, values of
parameters which characterize the family of TDOD curves, considered in
Subsection 3.2, are listed in Table \ref{t:RBgA}.
\begin{table}
\caption{Values of the cut, $\zeta_{\rm O}$,
the flow, $\kappa$, the fractional active gas mass,
$\mu_f$, the fractional long-lived star mass, $s_f$, the fractional
flowing gas mass, $D_f$,
the parameter, $c$, and the
normalization constant, $C_{\rm N}$, for
the family of theoretical differential oxygen
abundance distributions which contains the
curves plotted in Fig.\,\ref{f:pm11oht}, top left panel restricted
to A stage.   In addition, $\zeta_{\rm lin}=1.0012$ and $\zeta_\infty=1.1726$,
as listed in Table \ref{t:zita}.   For further details refer to the text.}
\label{t:RBgA}
\begin{center}
\begin{tabular}{lllllll} \hline
\multicolumn{1}{l}{$\zeta_{\rm O}$} &
\multicolumn{1}{c}{$-\kappa$} &
\multicolumn{1}{c}{$\mu_f$} &
\multicolumn{1}{c}{$s_f$} &
\multicolumn{1}{c}{$-D_f$} &
\multicolumn{1}{c}{$c$} &
\multicolumn{1}{c}{$C_{\rm N}$} \\
\hline
                  &             &             &             &             &                &             \\
0.00              & 1.2587E$+$0 & 1.7268E$+$0 & 2.8099E$+$0 & 3.5367E$+$0 & $+$1.2284E$+$0 & 5.3730E$-$2 \\
0.10              & 1.3860E$+$0 & 2.1334E$+$0 & 2.9360E$+$0 & 4.0694E$+$0 & $+$1.2176E$+$0 & 5.6624E$-$2 \\
0.20              & 1.5407E$+$0 & 2.6596E$+$0 & 3.0693E$+$0 & 4.7290E$+$0 & $+$1.2032E$+$0 & 5.9859E$-$2 \\
0.30              & 1.7323E$+$0 & 3.3487E$+$0 & 3.2072E$+$0 & 5.5560E$+$0 & $+$1.1839E$+$0 & 6.3482E$-$2 \\
0.35              & 1.8462E$+$0 & 3.7728E$+$0 & 3.2765E$+$0 & 6.0493E$+$0 & $+$1.1718E$+$0 & 6.5455E$-$2 \\
0.40              & 1.9754E$+$0 & 4.2627E$+$0 & 3.3450E$+$0 & 6.6077E$+$0 & $+$1.1574E$+$0 & 6.7546E$-$2 \\
0.50              & 2.2932E$+$0 & 5.4925E$+$0 & 3.4740E$+$0 & 7.9665E$+$0 & $+$1.1201E$+$0 & 7.2111E$-$2 \\
0.60              & 2.7752E$+$0 & 7.1749E$+$0 & 3.5792E$+$0 & 9.7541E$+$0 & $+$1.0654E$+$0 & 7.7246E$-$2 \\
0.65              & 3.0047E$+$0 & 8.2475E$+$0 & 3.6152E$+$0 & 1.0863E$+$1 & $+$1.0282E$+$0 & 8.0052E$-$2 \\
0.70              & 3.3446E$+$0 & 9.5204E$+$0 & 3.6341E$+$0 & 1.2154E$+$1 & $+$9.8155E$-$1 & 8.3030E$-$2 \\
0.75              & 3.7663E$+$0 & 1.1040E$+$1 & 3.6292E$+$0 & 1.3669E$+$1 & $+$9.2179E$-$1 & 8.6193E$-$2 \\
0.80              & 4.3029E$+$0 & 1.2839E$+$1 & 3.5919E$+$0 & 1.5456E$+$1 & $+$8.4346E$-$1 & 8.9553E$-$2 \\
0.90              & 5.9747E$+$0 & 1.7756E$+$1 & 3.3683E$+$0 & 2.0125E$+$1 & $+$5.8900E$-$1 & 9.6916E$-$2 \\
1.00              & 9.6022E$+$0 & 2.5143E$+$1 & 2.8067E$+$0 & 2.6950E$+$1 & $+$1.1200E$-$2 & 1.0523E$-$1 \\
$\zeta_{\rm lin}$ & 9.6714E$+$0 & 2.5253E$+$1 & 2.7969E$+$0 & 2.7050E$+$1 & \phantom{$+$}0 & 1.0534E$-$1 \\
1.10              & 2.3278E$+$1 & 3.6730E$+$1 & 1.6039E$+$0 & 3.7334E$+$1 & $-$2.2400E$+$0 & 1.1463E$-$1 \\
1.15              & 7.5661E$+$1 & 4.5018E$+$1 & 5.8956E$-$1 & 4.4607E$+$1 & $-$1.0964E$+$1 & 1.1977E$-$1 \\
1.16              & 1.3622E$+$2 & 4.6946E$+$1 & 3.3979E$-$1 & 4.6286E$+$1 & $-$2.1067E$+$1 & 1.2084E$-$1 \\
1.17              & 6.7115E$+$2 & 4.8979E$+$1 & 7.1595E$-$2 & 4.8051E$+$1 & $-$1.1033E$+$2 & 1.2192E$-$1 \\
$\zeta_\infty$    &  $+\infty$  & 4.9516E$+$1 & 0           & 4.8516E$+$1 & $-\infty$      & 1.2219E$-$1 \\
                                                                                   
\hline                         
\end{tabular}                     
\end{center}                      
\end{table}                       
The linear limit, the dominant flow limit, together with other curves
plotted in Fig.\,\ref{f:pm11oht}, top left panel, A stage, are also included.
Due to the occurrence of inflow regime, the flow, $\kappa$, and related
fractional flowing mass, $D$, are negative by definition.   All the parameters
listed in Table \ref{t:RBgA} exhibit a monotonic trend with the exception of
the long-lived star fractional mass, $s_f$, which attains a maximum at
$\zeta_{\rm O}\approx0.70$ and decreases to zero in the dominant flow
limit, $\zeta_{\rm O}=(\zeta_{\rm O})_\infty$.   In particular, the TDOD
fits to the EDOD within the range,
$0.35\appleq\zeta_{\rm O}\le(\zeta_{\rm O})_\infty$, related to $-1.85\appgeq
\kappa>-\infty$, $3.77\appleq\mu_f\appleq49.52$, $0\le s_f\appleq3.63$.   It
can also be seen the normalization constant, $C_{\rm N}$, is changed only
slightly with respect to the linear limit, $\zeta_{\rm O}=(\zeta_{\rm O})_
{\rm lin}$.

In the general case of a varying oxygen abundance within the flowing gas i.e.
varying cut, $(\zeta_{\rm O})_{\rm min}\le\zeta_{\rm O}\le(\zeta_{\rm O})_
{\rm max}$, the TDOD belonging to the family of curves passing through the
points, $(\phi_i,\psi_i)$, $(\phi_f,\psi_f)$, intersects all curves
characterized by constant $\zeta_{\rm O}$ in the range under consideration,
lying within the region bounded by the curves where $\zeta_{\rm O}=(\zeta_
{\rm O})_{\rm min}$ and $\zeta_{\rm O}=(\zeta_{\rm O})_{\rm max}$.   As shown
in Fig.\,\ref{f:pm11oht}, the TDOD fits to the EDOD over the whole range of
$\zeta_{\rm O}$, with the restriction, $(\zeta_{\rm O})_{\rm min}=0.35$, for
the A stage, top left panel (middle full curve).

Values of active gas, long-lived star and flowing gas mass fraction at the end
of each stage are listed in Table \ref{t:ralg} for the linear limit (upper
lines) and the lowest curve fitting to the data (lower lines) which
corresponds
to $(\zeta_{\rm O})_{\rm min}=0.35$ for the PM11n sample and $(\zeta_{\rm O})_
{\rm min}=0$ for Ra07n, PM11k, Ra07k samples, with regard to the initial A
stage.
\begin{table}
\caption{Values of active gas, long-lived star
and flowing gas mass fractions at the end of each
stage for the extreme situations of inflowing gas
during the earlier stage (A)
with oxygen abundance equal to $\zeta_{\rm O}=(\zeta_{\rm O})_{\rm lin}$
(upper lines) and $\zeta_{\rm O}=(\zeta_{\rm O})_{\rm min}$ (lower lines)
with respect to pre existing gas, in
connection with the thin (top cases) and thick (bottom cases) disk.
The values of the cut related to lower curves still consistent
with the data, plotted in Fig.\,\ref{f:pm11oht}, are
$(\zeta_{\rm O})_{\rm min}=0.35$ (PM11n sample), 0 (Ra07n, PM11k, Ra07k
samples).   During the subsequent F, C, and E stage, $\zeta_{\rm O}=(\zeta_
{\rm O})_{\rm lin}$ has been assumed.}
\label{t:ralg}
\begin{center}
\begin{tabular}{lllll} \hline
\multicolumn{1}{l}{U} &
\multicolumn{1}{c}{$(\mu_{\rm U})_f$} &
\multicolumn{1}{c}{$(s_{\rm U})_f$} &
\multicolumn{1}{c}{$(D_{\rm U})_f$} &
\multicolumn{1}{c}{sample} \\
\hline

      &             &             &                &       \\
A     & 2.5253E$+$1 & 2.7969E$+$0 & $-$2.7050E$+$1 & PM11n \\
      & 3.7728E$+$0 & 3.2765E$+$0 & $-$6.0493E$+$0 &       \\
F     & 2.7230E$+$1 & 1.5911E$+$1 & $-$4.2141E$+$1 &       \\
      & 4.0681E$+$0 & 5.2358E$+$0 & $-$8.3039E$+$0 &       \\
C     & 7.0432E$+$0 & 2.3164E$+$1 & $-$2.9207E$+$1 &       \\
      & 1.0523E$+$0 & 6.3194E$+$0 & $-$6.3717E$+$0 &       \\
E     & 3.3456E$-$1 & 2.4704E$+$1 & $-$2.4039E$+$1 &       \\
      & 4.9984E$-$2 & 6.5495E$+$0 & $-$5.5995E$+$0 &       \\
      &             &             &                &       \\
A     & 3.1135E$-$0 & 4.1536E$-$1 & $-$2.5289E$-$0 & Ra07n \\
      & 1.2393E$-$0 & 7.7920E$-$1 & $-$1.0185E$-$0 &       \\
F     & 1.6078E$-$0 & 1.8633E$-$0 & $-$2.4711E$-$0 &       \\
      & 6.3995E$-$1 & 1.3555E$-$0 & $-$9.9548E$-$1 &       \\
CE    & 7.5727E$-$2 & 2.4645E$-$0 & $-$1.5402E$-$0 &       \\
      & 3.0142E$-$2 & 1.5948E$-$0 & $-$6.2497E$-$1 &       \\
      &             &             &                &       \\
AC    & 1.8961E$-$0 & 1.3500E$-$0 & $-$2.2461E$-$0 & PM11k \\
      & 3.0081E$-$1 & 1.7505E$-$0 & $-$1.0513E$-$0 &       \\
CE    & 1.1633E$-$1 & 1.9835E$-$0 & $-$1.0999E$-$0 &       \\
      & 1.8456E$-$2 & 1.8510E$-$0 & $-$8.6944E$-$1 &       \\
      &             &             &                &       \\
A     & 4.5570E$-$0 & 8.1266E$-$1 & $-$4.3696E$-$0 & Ra07k \\
      & 1.3223E$-$0 & 1.0925E$-$0 & $-$1.4148E$-$0 &       \\
F     & 4.2824E$-$0 & 2.6077E$-$0 & $-$5.8901E$-$0 &       \\
      & 1.2474E$-$0 & 1.5537E$-$0 & $-$1.8011E$-$0 &       \\
CE    & 6.8384E$-$1 & 4.4615E$-$0 & $-$4.1454E$-$0 &       \\
      & 1.9836E$-$1 & 2.0941E$-$0 & $-$1.2925E$-$0 &       \\
\hline                            
\end{tabular}                     
\end{center}                      
\end{table}                       
The linear limit is assumed during the following F, C and E stage.
An inspection of Table \ref{t:ralg} discloses that the active gas mass
fraction is lowered by a factor up to about 7, while long-lived star and
flowing gas mass fraction are lowered by a factor up to about 4, passing from
the linear limit, $\zeta_{\rm O}=(\zeta_{\rm O})_{\rm lin}$, to the lowest
curve fitting to the data, $\zeta_{\rm O}=(\zeta_{\rm O})_{\rm min}$, at the
end of evolution.

\section{Discussion} \label{disc}

As shown in Fig.\,\ref{f:pm11ohl}, the EDOD inferred from PM11 and Ra07
samples for the thin and the thick disk (see Table \ref{t:samp}) can be
fitted by a
straight line on four, three or two adjacent regions.   In the light of simple
MCBR
models of chemical evolution, the TDOD is represented by a family of curves
(including the straight line) passing through the starting and the ending
point, $(\phi_i,\psi_i)$, $(\phi_f,\psi_f)$, respectively, of a selected stage
of evolution.   Accordingly, different stages of evolution relate to different
linear fits to the EDOD, where an initial strong inflow regime $(a>0)$ is
followed by a steady $(a=0)$ or moderate $(-0.42<a<0)$ inflow regime and
finally by a stagnation $(a=-0.42)$ or outflow $(a<-0.42)$ regime.   For
further details refer to an earlier attempt (C11).

The above description is in agreement with the results of hydrodynamical
simulations, where quasi equilibrium is attained between inflowing gas,
outflowing gas and gas lost via star formation, after an early stage of strong
inflow and, presumably, before a late stage of strong outflow (e.g., Finlator
and Dav\'e 2008; Dav\'e et al. 2011a,b; 2012).   Within the formalism of
simple MCBR models, both the inflow and the outflow rate are proportional to
the star formation rate and, for this reason, only the net effect, $\kappa=
\kappa_{\rm ifl}+\kappa_{\rm ofl}$, $\zeta_{\rm O}=(\zeta_{\rm O})_{\rm ifl}+
(\zeta_{\rm O})_{\rm ofl}$, is considered.

Though the assumption of null stellar mass fraction holds to an acceptable
extent for the starting configuration, still the presence of stars with lower
oxygen abundance than in sample objects, $\phi<0.4$ or [O/H$]<-0.4$ for the
thin disk and $\phi<0.2$ or [O/H$]<-0.7$ for the thick disk, has
to be considered.   To this aim, samples unbiased towards low oxygen abundance
(not available at present) should be used.

The thick disk is expected to be
affected from this inconvenient to a larger extent with respect to the thin
disk, due to a more extended low-metallicity tail.   In any case, the results
of the current paper may be thought of as representative of the chemical
evolution of the thin disk for $\phi\ge0.4$ or [O/H$]\ge-0.4$ and of the thick
disk for $\phi\ge0.3$ or [O/H$]\ge-0.5$, the last related to [Fe/H$]\ge-1$,
where the data are unbiased (Ramirez et al. 2007; Petigura and Marcy 2011).

As shown in Fig.\,\ref{f:pm11oht}, the TDOD fits to the EDOD regardless of the
value of the cut, $\zeta_{\rm O}$ i.e. the oxygen abundance ratio of flowing
to pre existing gas, within a plausible range where the physical meaning of
the problem is preserved.   The sole exception relates to the PM11n sample
(Fig.\,\ref{f:pm11oht}, top left), A stage, where curves within the range,
$0\le\zeta_{\rm O}<0.35$, provide a bad fit to the data, which implies inflow
of low-metallicity gas within the thin disk could not take place when the pre
existing gas was sufficiently oxygen-enriched, [O/H$]\appgeq-0.4$.

The above result can be extended to the case of variable cut, $(\zeta_{\rm O})
_{\rm min}\le\zeta_{\rm O}\le(\zeta_{\rm O})_{\rm max}$, where the
TDOD lies between the region bounded by the curves related to $(\zeta_{\rm O})
_{\rm min}$ and $(\zeta_{\rm O})_{\rm max}$, respectively.   Then the
differential distribution of oxygen abundance in thin disk and thick disk
stars (leaving aside the low-metallicity tail) depends only slightly on the
oxygen abundance of the flowing gas.

Biases towards low-metallicity and high-metallicity stars may be corrected
under the assumption that the TDOD related to $\phi<\phi_i$ and $\phi>\phi_f$,
respectively, allow analytical continuation.   Let $\phi_{\rm min}$ and
$\phi_{\rm max}$ be the unknown initial and final normalized oxygen abundance
related to stages A and E, respectively.

The fractional number of long-lived stars with normalized oxygen abundance,
$\phi_{\rm min}\le\phi\le\phi_i$, can be inferred from Eq.\,(\ref{eq:ptin})
as:
\begin{lefteqnarray}
\label{eq:dNA}
&& \frac{(\delta N)_{\rm A}}{N_f-N_i}=-10^{b_{\rm A}}\frac{\hat{p}^\pprime}
{(Z_{\rm O})_\odot} \nonumber \\
&& \phantom{\frac{(\delta N)_{\rm A}}{N_f-N_i}=}\times
\left[(1-c_{\rm A}\phi_i)^{(Z_{\rm O})_\odot/(c_{\rm A}
\hat{p}^\pprime)}-(1-c_{\rm A}\phi_{\rm min})^{(Z_{\rm O})_\odot/(c_{\rm A}
\hat{p}^\pprime)}\right]~~;
\end{lefteqnarray}
or, in the linear limit $(c\to0)$, from Eq.\,(\ref{eq:ptil}) as:
\begin{lefteqnarray}
\label{eq:dNAl}
&& \frac{(\delta N)_{\rm A}}{N_f-N_i}=\frac1{\ln10}\frac1{a_{\rm A}}\left[
\exp_{10}(a_{\rm A}\phi_i+b_{\rm A})-\exp_{10}(a_{\rm A}\phi_{\rm min}+
b_{\rm A})\right]~~;
\end{lefteqnarray}
where, in any case, the maximum value relates to $\phi_{\rm min}=0$.

The fractional number of long-lived stars with normalized oxygen abundance,
$\phi_f\le\phi\le\phi_{\rm max}$, can be inferred from Eq.\,(\ref{eq:ptin})
as:
\begin{lefteqnarray}
\label{eq:dNE}
&& \frac{(\delta N)_{\rm E}}{N_f-N_i}=-10^{b_{\rm E}}\frac{\hat{p}^\pprime}
{(Z_{\rm O})_\odot} \nonumber \\
&& \phantom{\frac{(\delta N)_{\rm E}}{N_f-N_i}=}\times
\left[(1-c_{\rm E}\phi_{\rm max})^{(Z_{\rm O})_\odot/(c_
{\rm E}\hat{p}^\pprime)}-(1-c_{\rm E}\phi_f)^{(Z_{\rm O})_\odot/(c_{\rm E}
\hat{p}^\pprime)}\right]~~;
\end{lefteqnarray}
or, in the linear limit $(c\to0)$, from Eq.\,(\ref{eq:ptil}) as:
\begin{lefteqnarray}
\label{eq:dNEl}
&& \frac{(\delta N)_{\rm E}}{N_f-N_i}=\frac1{\ln10}\frac1{a_{\rm E}}\left[
\exp_{10}(a_{\rm E}\phi_{\rm max}+b_{\rm E})-\exp_{10}(a_{\rm E}\phi_f+
b_{\rm E})\right]~~;
\end{lefteqnarray}
where, in any case, the maximum value relates to $\phi_{\rm max}=A_{\rm O}/
(Z_{\rm O})_\odot$ via Eq.\,(\ref{eq:AO}), which implies $Z=1$ and may be
replaced, with unrelevant numerical change in the results, by $\phi_{\rm max}
\to+\infty$.

The fractional star mass change at the end of evolution, due to the presence
of (undetected) low-metallicity and high-metallicity tail in the EDOD, can be
inferred under the assumption that star number is proportional to star mass,
$\delta N/N=\delta s_f/s_f$.   The result is:
\begin{lefteqnarray}
\label{eq:dsf}
&& \frac{\delta s_f}{s_f}=\frac{(\delta N)_{\rm A}+(\delta N)_{\rm E}}
{N_f-N_i}~~;
\end{lefteqnarray}
where the values of the parameters can be deduced from the linear fits to the
EDOD, as shown in Tables \ref{t:abs} and \ref{t:figs}.   The results are
listed in Table \ref{t:code}, with regard to upper values, in the linear
limit.
\begin{table}
\caption{Upper values of the fractional star mass change due to the
low-metallicity tail (A), the high-metallicity tail (E), both
tails (A+E), under the assumption that the linear fit to the EDOD maintains
unchanged outside the oxygen abundance range exhibited by the parent sample.
Calculations were performed in the linear limit using values listed in Tables
\ref{t:abs} and \ref{t:figs} together with $\phi_{\rm min}=0$ and $\phi_
{\rm max}=A_{\rm O}/(Z_{\rm O})_\odot$ i.e. $Z=1$ which, for practical
purposes, is equivalent to $\phi_{\rm max}\to+\infty$.   See text for further
details.}
\label{t:code}
\begin{center}
\begin{tabular}{llll} \hline
\multicolumn{1}{l}{$(\delta s_f)_{\rm A}/s_f$} &
\multicolumn{1}{c}{$(\delta s_f)_{\rm E}/s_f$} &
\multicolumn{1}{c}{$(\delta s_f)_{\rm A+E}/s_f$} &
\multicolumn{1}{c}{sample} \\
\hline
             &              &              &       \\
4.1925 E$-$3 & 2.8929 E$-$3 & 7.0854 E$-$3 & PM11n \\ 
6.6414 E$-$2 & 1.1666 E$-$2 & 7.8079 E$-$2 & Ra07n \\
1.8307 E$-$1 & 1.8189 E$-$2 & 2.0126 E$-$1 & PM11k \\
3.1718 E$-$2 & 7.4427 E$-$2 & 1.0614 E$-$1 & Ra07k \\
\hline       
\end{tabular}
\end{center} 
\end{table}  

It can be seen the larger contribution arises from the low-metallicity tail
for PM11k sample and from the high-metallicity tail for Ra07k sample.   The
reason is that, in both cases, the paucity of data makes more extended
metallicity tails, as shown in Fig.\,\ref{f:pm11ohl}.   The total change in 
fractional star mass does not exceed
about 10\% at most, with the exception of the (poorly populated) PM11k sample,
where it raises to about 20\%.   Accordingly, the results of the current paper
hold to an acceptable extent even if the linear trend shown by the EDOD is
extended to undetected metallicity tails.

In general, the disk is usually conceived as made of
two main subsystems: the thick disk and the thin disk. Accordingly,
the EDOD related to the disk depends
on the thick to thin disk mass ratio, $M_{\rm K}/M_{\rm N}$,
which is poorly known at present. Values already quoted in
literature span a wide range, from some percent (e.g., Holmberg
et al. 2007) to about unity (e.g., Fuhrmann 2008), or
even indeterminate in the sense that no distinction can be
made (e.g., Norris 1987; Ivezic et al. 2008; Bovy et al. 2012).   Very low
values of $M_{\rm K}/M_{\rm N}$ could be biased unless corrected taking into
consideration the different heights of the two subsystems on the Galactic
plane at the Sun.

The idea of a thick disk - thin disk collapse is not in contradiction with
related specific angular momentum distribution, as shown in earlier attempts
(Wyse and Gilmore 1992; Ibata and Gilmore 1995).   If it is the case, in the
light of the model, the thick and thin disk gas and star fractional mass are
normalized to the same initial mass and the mass ratio of the thick to the
thin disk at the end of evolution reads:
\begin{equation}
\label{eq:MKN}
\frac{M_{\rm K}}{M_{\rm N}}=\frac{(s_f)_{\rm K}}{(\mu_f)_{\rm K}+
(\mu_f)_{\rm N}+(s_f)_{\rm N}}~~;
\end{equation}
where $s_f$ and $\mu_f$ relate to the end of evolution and the indexes, K and
N, denote the thick and the thin disk, respectively.   From the results listed
in Table \ref{t:ralm}, it can be inferred $M_{\rm K}/M_{\rm N}=0.08,$ 1.89,
for PM11 and Ra07 samples, respectively.    The former value would argue for
a thick disk - thin disk collapse, but the PM11k sample is poorly populated to
draw firm conclusions.   On the other hand, the latter value would be against
a thick disk - thin disk collapse, unless a massive thick disk (with respect
to the thin disk) is recognized.    
Finally, if a single (thick + thin) disk population exists, the EDOD maintains 
close to its counterpart inferred both from the PM11n sample, due to the
smallness of the PM11k sample, and from the Ra07n sample, as shown in
an earlier attempt (Caimmi and Milanese 2009).

The family of TDOD curves, considered for each stage of evolution, is
characterized by two common points, $(\phi_i,\psi_i)$, $(\phi_f,\psi_f)$,
which make the boundary of the related linear fit.   As shown in Table
\ref{t:RBgA} for the PM11n sample, A stage, curves with increasing cut,
$\zeta_{\rm O}$, exhibit a monotonic trend for the remaining parameters,
with the exception of the star mass fraction at the end of evolution, $s_f$,
which attains a maximum.   Accordingly, for a selected stage of evolution and
fixed starting and ending point of the related linear fit to the EDOD, the
star formation efficiency cannot exceed a threshold.   In the case under
consideration, the model predicts an amount not exceeding 3.63 the initial
mass of the thin disk was turned into stars during the earlier (A) stage of
evolution.

\section{Conclusion}
\label{conc}

The main results of the current paper may be summarized as follows.
\begin{description}
\item[(1)\hspace{2.0mm}]
The empirical differential oxygen abundance distribution (EDOD)
inferred from two different samples for both the thin and the
thick disk is consistent with a linear trend within four, three or two
regions (Fig.\,\ref{f:pm11ohl}).
\item[(2)\hspace{2.0mm}]
A family of theoretical differential oxygen abundance distribution 
\linebreak
(TDOD) curves, passing through the starting point, $(\phi_i,\psi_i)$,
and the ending point, $(\phi_f,\psi_f)$, of the linear fit to the
EDOD, is defined within the framework of simple multistage closed
(box+reservoir) (MCBR) models.   In addition, regions where the
EDOD exhibits a linear trend are related to stages of evolution
characterized by different inflow or outflow rate and/or different
oxygen abundance within the flowing gas.   The extreme curves
of the family on the $({\sf O}\phi\psi)$ plane are related to
$\zeta_{\rm O}=0,\,(\zeta_{\rm O})_\infty$, the last 
corresponding to $\kappa\to-\infty$, for inflowing gas and
$\zeta_{\rm O}=(\zeta_{\rm O})_\infty,\,A_{\rm O}/(Z_{\rm O})_f$,
the first  corresponding to $\kappa\to+\infty$, for outflowing gas.
\item[(3)\hspace{2.0mm}]
For a family of TDOD curves related to an assigned stage of evolution, all the
parameters show a monotonic trend passing from the lower to the upper curve,
with the exception of the fractional stellar mass at the end of evolution,
which attains a maximum and, in consequence, cannot exceed a threshold.   In
particular, the model predicts an amount not exceeding 3.63 the initial mass
(Table \ref{t:RBgA}) of the thin disk is turned into stars during the earlier
(A) stage of evolution, in connection with the EDOD inferred from the PM11n
sample.
\item[(4)\hspace{2.0mm}]
The special case of steady inflow regime, where the TDOD reduces to a
horizontal line, is consistent with the results of hydrodynamical simulations,
where quasi equilibrium is attained between inflowing gas, outflowing gas and
gas lost via star formation (e.g., Finlator and Dav\'e 2008; Dav\'e et al.
2011a,b; 2012).
\item[(5)\hspace{2.0mm}]
If the linear trend inferred for the EDOD extends towards undetected low and
high oxygen abundance, then the fractional star mass change does not exceed
about 10\% for well populated samples (PM11n, Ra11n, Ra11k) and about 20\% for
poorly populated samples (PM11k).
\item[(6)\hspace{2.0mm}]
Under the assumption of a thick disk - thin disk collapse, model predictions
yield a mass ratio, $M_{\rm K}/M_{\rm N}=0.08,$ 1.89, with regard to PM11 and
Ra07 samples, respectively.   The latter alternative is in contradiction with
current observations, which seem to exclude the presence of a massive thick
disk, while the former remains available on this respect.   If thick disk and
thin disk stars belong to a single population (e.g., Bovy et al. 2012), no
prediction can be made for or against a thick disk - thin disk collapse and
the TDOD is expected to be slightly different from its counterpart related to
the thin disk (Caimmi and Milanese 2009).
\end{description}
The main uncertainties on the EDOD used in the current paper are related to
(i) poorly populated samples (PM11k); (ii) biased samples (all) towards low
metallicities, [Fe/H$]<-1.0$ or [O/H$]<-0.5$, which are expected to affect
mainly the thick disk, where the overall metallicity or oxygen abundance are
lower than in the thin disk, even if the low-metallicity tail is neglected.
On the other hand, the prediction of early strong inflow implies the main
features of thick and thin disk evolution are captured by the EDOD inferred
from PM11n, Ra07n, Ra07k samples, even if the low-metallitity tail is
neglected.

\section*{Acknowledgements}
Thanks are due to the referee, S. Ninkovi\'c, for useful comments.

\end{document}